\documentclass[12pt,a4paper]{article}
\usepackage[latin1]{inputenc}
\usepackage[english]{babel}
\usepackage[T1]{fontenc}
\usepackage{ae,aecompl}
\usepackage{amsmath,amssymb,amsfonts}
\usepackage{bm}
\usepackage{ifpdf}
\ifpdf
  \usepackage[pdftex,a4paper,hmargin=25mm,vmargin=25mm]{geometry}
\else
  \usepackage[dvips,a4paper,hmargin=25mm,vmargin=25mm]{geometry}
\fi

\numberwithin{equation}{section}

\DeclareTextCompositeCommand{\v}{PD1}{r}{r}

\begin{document}

\newcommand{\email}[1]{Electronic address: \href{mailto:#1}{#1}}
\newcommand{\HL}{Ho\v{r}ava-Lifshitz }

\newcommand{\be}{\begin{equation}}
\newcommand{\ee}{\end{equation}}
\newcommand{\bea}{\begin{eqnarray}}
\newcommand{\eea}{\end{eqnarray}}
\newcommand{\nn}{\nonumber\\}

\newcommand{\mc}[1]{\mathcal{#1}}
\newcommand{\mr}[1]{\mathrm{#1}}
\newcommand{\rd}{\mathrm{d}}
\newcommand{\dx}[1]{\mathrm{d}x^{#1}}
\newcommand{\p}{\partial}
\newcommand{\vect}[1]{\bm{#1}}

% space-time metric and curvature (4-dimensional)
\newcommand{\fg}{{}^{4}\!g}
\newcommand{\fR}{{}^{4}\!R}
\newcommand{\fnabla}{\nabla^{(4)}}
\newcommand{\tildeR}{{}^{4}\!\tilde{R}}
\newcommand{\cLR}[1][g_{ij}]{\mathcal{L}_R(#1)}

\newcommand{\cG}{\mathcal{G}} % supermetric

\begin{center}
{\Large Hamiltonian analysis of non-projectable modified \\
\vspace{.3em}
$F(R)$ Ho\v{r}ava-Lifshitz gravity}\\
\vspace{1em}
Masud Chaichian$^{a,b}$,\footnote{\email{masud.chaichian@helsinki.fi}} 
Markku Oksanen$^a$,\footnote{\email{markku.oksanen@helsinki.fi}} and 
Anca Tureanu$^{a,b}$\footnote{\email{anca.tureanu@helsinki.fi}}\\
\vspace{1em}
\textit{${}^a$Department of Physics, University of Helsinki, P.O. Box
64,\\ FI-00014 
Helsinki, Finland\\
${}^b$Helsinki Institute of Physics, P.O. Box 64, FI-00014 Helsinki,
Finland}
\end{center}

\begin{abstract}
We study a version of the recently proposed modified $F(R)$
Ho\v{r}ava-Lifshitz gravity that abandons the projectability condition
of the lapse variable. We discovered that the projectable version of
this theory has a consistent Hamiltonian structure, and that the theory
has interesting cosmological solutions which can describe the eras of
accelerated expansion of the universe in a unified manner. The usual
Ho\v{r}ava-Lifshitz gravity is a special case of our theory. Hamiltonian
analysis of the non-projectable theory, however, shows that this theory
has serious problems. These problems are compared with those found in
the original Ho\v{r}ava-Lifshitz gravity. A general observation on the
structure of the Poisson bracket of Hamiltonian constraints in all
theories of the Ho\v{r}ava-Lifshitz type is made: in the resulting
tertiary constraint the highest order spatial derivative of the lapse
$N$ is always of uneven order. Since the vanishing of the lapse ($N=0$)
is required by the preservation of the Hamiltonian 
constraints under time evolution, we conclude that the non-projectable
version of the theory is physically inconsistent.
\end{abstract}

\section{Introduction}

Last year the so-called \HL theory of gravity was proposed
\cite{Horava:2009uw} (see also \cite{Horava:2008ih,Horava:2009if}). This
theory is a candidate for a quantum field theory of gravity that aims to
provide an ultraviolet (UV) completion of General Relativity (GR). At
short distances it describes interacting nonrelativistic gravitons. \HL
gravity exhibits anisotropic scaling of space and time coordinates
\be \label{anisotropic_scaling}
\vect{x} \rightarrow b \vect{x} \, ,\quad t \rightarrow b^z t
\ee
with a dynamic critical exponent $z=1,2,3,\ldots$. In the UV regime the
value of the critical exponent $z$ is chosen so that the gravitational
coupling constant $\kappa^2$ is dimensionless. In $(D+1)$-dimensional
space-time we have the scaling dimension $[\kappa^2]=z-D$. Thus the
choice $z=D$ is argued to ensure that the theory is power-counting
renormalizable. For the usual case of 3-dimensional space, $D=3$, we
choose $z=3$.

The space-time manifold $\mc{M}$ is assumed to possess a foliation
structure that enables one to define $\mc{M}$ as a union of space-like
hypersurfaces $\Sigma_t$ of constant time $t$. Due to the foliation the
space-time is invariant under the foliation-preserving diffeomorphisms,
whose infinitesimal generators are of the form
\be
\delta\vect{x} = \zeta(t, \vect{x}) \,,\quad \delta t = f(t) \,,
\ee
instead of the full diffeomorphism invariance of GR. For simplicity the
topological structure of space-time is assumed to be such that every
leaf $\Sigma_t$ of the foliation is equivalent to a fixed manifold
$\Sigma$: $\mc{M} \cong \mathbb{R} \times \Sigma$. The preferred
foliation of $\mc{M}$ enables the inclusion of spatial covariant
derivatives into the action, which improve the UV behaviour, while
avoiding time derivatives higher than the second order, which are known
to produce problematic ghosts.

At low energies and large distances the critical exponent is expected to
flow to $z=1$, so that the theory can coincide with GR. The Lorentz
symmetry emerges at low energies as an accidental or approximate
symmetry, but it is absent in the fundamental description.

The theory comes in two flavors, with or without the projectability
condition that requires the lapse to depend only on the time coordinate,
$N=N(t)$. The projectability condition is one of the features that makes
the theory differ from GR. Note, however, that many solutions of GR, as
well as of non-projectable \HL gravity, respect the condition $N=N(t)$
even thought the general theory does not.

In the original theory an additional symmetry, the condition of detailed
balance, is assumed. It defines the potential part of the gravitational
action in terms of a variation of a $D$-dimensional action on the
spatial hypersurface with respect to the spatial metric. The purpose of
the detailed balance condition is to reduce the number of independent
couplings to 3 --- otherwise there are 9 independent couplings. We,
however, do not  assume this condition, so all terms that have
appropriate scaling properties and that are covariant under
foliation-preserving diffeomorphisms can be included.

This theory has received a lot of attention and many potential problems
have been discovered. Some of the problems are very serious. First it
was found that GR is not recovered at large distances if the detailed
balance condition is assumed \cite{Lu:2009em,Kehagias:2009is}. A
``phenomenologically viable'' version of the theory without the detailed
balance condition was soon introduced \cite{Sotiriou:2009}. Due to the
reduced diffeomorphism symmetry group there is an additional ``half''
scalar degree of freedom in \HL gravity. It has been shown to be
strongly coupled at all scales by considering perturbations about a
reasonable vacuum \cite{Charmousis:2009}, regardless whether the
detailed balance is assumed or not. This suggests that perturbative GR
cannot be reproduced in \HL gravity \cite{Charmousis:2009}, and that the
theory could be ruled out by existing observations on the gravitational
radiation of binary pulsars, which agree with linearized GR. The
low-energy regime of the theory was further analyzed 
in Ref.~\cite{Blas:2009,Kobakhidze:2009} where problems with instability
and strong coupling of the extra degree of freedom were found. Since
then these problems have been confirmed in various papers.
Although the non-projectable version of the theory may not give GR as
the limit at large distances, some other scenarios, such as the
chameleon, may solve this problem.
A ``healthy extension'' of \HL gravity was proposed in
Ref.~\cite{Blas:2010} that is argued to be free from at least some of
the pathologies of the original theory, since the extra scalar mode has
a healthy quadratic action. This is achieved by adding terms that
involve the spatial 3-vector $N^{-1}\nabla_i N$ into the action. We
assume that the Hamiltonian takes the canonical form [see \eqref{Ha}]
which rules out such terms. Hamiltonian formalism of the healthy
extension has been studied in Ref.~\cite{Kluson:2010nf}.

Renormalizability of the \HL gravity has been investigated beyond the
power-counting scheme in Ref.~\cite{Orlando:2009}.

When the projectability condition is assumed, the theory has a quite
simple and consistent Hamiltonian structure. The algebra of constraints
was shown to be closed for $z=1,2$ in Ref.~\cite{Horava:2008ih}, and
this holds for a higher scaling exponent $z$ as well. The Hamiltonian
structure of \HL gravity without the projectability condition has been
analyzed particularly in Ref.~\cite{Li:2009,Henneaux:2010}. The
non-projectable theory is physically inconsistent for generic couplings
\cite{Henneaux:2010}, including the case with detailed balance
\cite{Li:2009}. In the case of low-energy effective action a consistent
set of constraints can be obtained by imposing an additional constraint
($\pi=0$) \cite{Henneaux:2010,Pons:2010,Restuccia:2010}.

Recently we proposed the modified $F(R)$ \HL gravity
\cite{Carloni:2010,Chaichian:2010yi} that combines the interesting
cosmological aspects of $F(R)$ gravity and the possible UV finiteness of
\HL gravity. In particular, we demonstrated that the solution of
spatially-flat FRW equation has two branches: one that coincides with
the usual $F(R)$ gravity for a certain choice of parameters, and one
that is totally new and typical only for \HL gravity. It was shown that
unlike to standard \HL gravity, our $F(R)$ \HL gravity enables the
possibility to unify the early-time inflation with the late-time
acceleration in accord with the scenario of Ref.~\cite{Nojiri:2003ft}. 
 In this paper we present the Hamiltonian analysis of the
non-projectable version of this theory, where the lapse $N$ depends also
on the spatial coordinates: $N=N(t,\vect{x})$. Expectedly the
Hamiltonian structure of this theory turns out to be more complex than
in the projectable case.
Our analysis should be of interest to everyone interested in the
Hamiltonian formalism of gravity, and of modified gravity in particular.

\section{Hamiltonian analysis}

\subsection{Action}

We assume the ADM decomposition of space-time \cite{Arnowitt:1962hi} 
(for reviews and mathematical background, see \cite{ADMreviewmath}). 
The metric tensor of space-time is decomposed in terms of the ADM
variables as
\be \label{spacetime_metric}
\fg_{\mu\nu} \dx{\mu} \dx{\nu} = - (N^2 - N_i N^i) \rd t^2 + N_i (\rd t
\dx{i} + \dx{i} \rd t) + g_{ij} \dx{i} \dx{j} \,,
\ee
where $N$ is the lapse, $N^i$ is the shift vector, $g_{ij}$ is the
spatial metric tensor, and $x^i, i = 1, 2, 3$ are spatial coordinates on
the $t=\mr{constant}$ hypersurface $\Sigma_t$. 
The covariant derivatives defined by the metric tensors $\fg_{\mu\nu}$
and $g_{ij}$ are denoted by $\fnabla_\mu$ and $\nabla_i$, respectively. 
The extrinsic curvature of the hypersurface $\Sigma_t$ is
\be
K_{ij} = \frac{1}{2N} \left( \dot{g}_{ij} - 2\nabla_{(i} N_{j)} \right)
\,,
\ee
where the dot denotes the derivative with respect to time $t$. The
scalar 
associated to the extrinsic curvature is denoted by $K=g^{ij}K_{ij}$. 
The (intrinsic) curvature of the space $\Sigma_t$ is defined by the 
spatial metric $g_{ij}$ in the usual manner. The natural 
invariant volume element of space-time is decomposed
\be
\rd^4 x \sqrt{-\fg} = \rd t \rd^3\vect{x} \sqrt{g}N \, .
\ee

The action of the non-projectable version of the modified $F(R)$ \HL
gravity is defined similarly as in the projectable case
\cite{Carloni:2010}:
\bea \label{HLFR_action}
&& S_F = \frac{1}{\kappa^2}\int\rd t \rd^3\vect{x} \sqrt{g} N
F\left(\tildeR\right)\, ,\nn
&& \tildeR \equiv K_{ij} K^{ij} - \lambda K^2 +
2 \mu \fnabla_\mu \left( n^\mu \fnabla_\nu n^\nu - n^\nu \fnabla_\nu
n^\mu \right) - \cLR \, .
\eea
Here $\lambda$ and $\mu$ are constants, $n^\mu$ is the unit normal 
to the spatial hypersurfaces $\Sigma_t$, and $\cLR$
is a function of the three-dimensional metric $g_{ij}$ and the
covariant derivatives $\nabla_i$ defined by this metric.
The crucial difference compared to the theory we proposed and 
analyzed in Ref.~\cite{Carloni:2010} is that the lapse function $N$
does 
not obey the projectability condition, i.e. it depends also on the
spatial 
coordinates, $N=N(t,\vect{x})$.
For the Hamiltonian analysis of modified Ho\v{r}ava-Lifshitz-like $F(R)$
gravity, which is a special case of the general projectable theory
\cite{Carloni:2010}, 
one can see Ref.~\cite{Chaichian:2010yi,Kluson:2010xx}. This special 
case with the further restriction to the parameter value $\mu=0$ has 
been proposed and analyzed in Ref.~\cite{Kluson:2009xx}. Yet another
special case, given by $F(\tildeR)=\tildeR$ and $\cLR=-f(R)$, has been
studied in Ref.~\cite{Wang:2010}.

By introducing two auxiliary fields $A$ and $B$ we can write the
action \eqref{HLFR_action} into a form that is linear in $\tildeR$:
\be \label{action_aux}
S_F = \frac{1}{\kappa^2}\int\rd t \rd^3\vect{x} \sqrt{g} N \left[ B
\left( \tildeR - A \right) +
F(A) \right] \, .
\ee
The variation with respect
to $B$ yields $\tildeR = A$, which can be inserted back into the
action \eqref{action_aux} in order to produce the original action
\eqref{HLFR_action}. The variation with respect to $A$ yields $B =
F'(A)$,
where $F'$ denotes the derivative of $F$ with respect to its
argument. Thus \eqref{action_aux} reduces to the action
\eqref{HLFR_action} 
when these equations of motion are imposed.

First we rewrite $\tildeR$ in \eqref{action_aux} in a more
explicit and useful form (see \eqref{HLFR_action} for the definition of
$\tildeR$). The unit normal $n^\mu$ to the hypersurface $\Sigma_t$
in space-time can be written in terms of the lapse and the shift
vector as $n^\mu = (n^0, n^i) = \left(\frac{1}{N}, -
\frac{N^i}{N}\right)$. The corresponding one-form is $n_\mu =
-N\fnabla_\mu t = (-N, 0, 0, 0)$. The term in \eqref{HLFR_action} that
involves the unit normal can be written as
\be \label{mu-term}
\fnabla_\mu \left( n^\mu \fnabla_\nu n^\nu - n^\nu \fnabla_\nu n^\mu
\right) 
= \fnabla_\mu \left( n^\mu K \right) - \frac{1}{N} \Delta N \, ,
\ee
where the spatial Laplacian is $\Delta = g^{ij} \nabla_i \nabla_j$.
Thus we can rewrite $\tildeR$
\be \label{tildeR_2nd}
\tildeR = K_{ij} \cG^{ijkl} K_{kl} + 2\mu \fnabla_\mu \left(
n^\mu K \right) - \frac{2\mu}{N} \Delta N - \cLR \,,
\ee
where the ``generalized De~Witt metric'' was introduced
\be \label{cG}
\cG^{ijkl} = \frac{1}{2}\left(
g^{ik} g^{jl} + g^{il} g^{jk} \right) - \lambda
g^{ij} g^{kl} \, .
\ee
Introducing \eqref{tildeR_2nd} into \eqref{action_aux} and
performing integrations by parts yields the action
\bea \label{HLFR_action_final}
S_F &=& \frac{1}{\kappa^2}\int\rd t \rd^3 \vect{x} \sqrt{g}
\Bigl\{ N \left[ B \left( K_{ij} \cG^{ijkl} K_{kl} - \cLR - A \right) +
F(A) \right] \nn
&&\qquad\qquad\qquad \left.  - 2\mu K \left(  \dot{B} - N^i \p_i B
\right)
 - 2\mu N \Delta B \right\} \, ,
\eea
where the integral is taken over the union $\mc{U}$ of the
$t=\rm{constant}$ hypersurfaces $\Sigma_t$ with $t$ over
some interval in $\mathbb{R}$, and we have written
$N n^\mu \fnabla_\mu B = \dot{B} - N^i \p_i B$.
We assume that the boundary integrals over $\p\mc{U}$ and
$\p\Sigma_t$ vanish.

\subsection{Hamiltonian and momentum constraints}

In the Hamiltonian formalism the field variables
$g_{ij}$, $N$, $N^i$, $A$ and $B$ have the canonically conjugated
momenta $\pi^{ij}$, $\pi_N$, $\pi_i$, $\pi_A$ and $\pi_B$,
respectively. Because the action does not depend on the time derivative 
of $N$, $N^i$ or $A$, their conjugated momenta are the primary
constraints:
\be \label{p_constraints}
\pi_N(\vect{x}) \approx 0\, ,\quad \pi_i(\vect{x}) \approx 0\, ,
\quad \pi_A(\vect{x}) \approx 0\, .
\ee
For the spatial metric and the field $B$ we have
the momenta
\bea
\pi^{ij} &=& \frac{\delta S_F}{\delta \dot{g}_{ij}}
= \frac{1}{\kappa^2}\sqrt{g} \left[ B \mc{G}^{ijkl} K_{kl} -
\frac{\mu}{N} g^{ij}
\left(  \dot{B} - N^k \p_k B \right) \right]\, ,\label{pi^ij}\\
\pi_B &=& \frac{\delta S_F}{\delta \dot{B}}
= - \frac{2\mu}{\kappa^2} \sqrt{g} K \, .\label{pi_B}
\eea
We assume $\mu\neq 0$ so that the momentum \eqref{pi_B} does not
vanish. 
When $\lambda\neq 1/3$, the generalized De~Witt metric \eqref{cG} has 
the inverse
\be \label{inversecG}
\cG_{ijkl} = \frac{1}{2}\left( g_{ik} g_{jl} + g_{il} g_{jk} \right)
- \frac{\lambda}{3\lambda - 1} g_{ij} g_{kl}\, ,\quad 
 \cG_{ijkl} \cG^{klmn} = \delta_{(i}^{(m} \delta_{j)}^{n)} \, .
\ee
However, as long as $\mu\neq 0$, the invertibility of \eqref{cG} is not
that significant in our theory, because $K$ is given by \eqref{pi_B}, $K
= - \frac{\kappa^2}{2\mu\sqrt{g}} \pi_B$, and therefore we have
\be
\cG^{ijkl} K_{kl} = K^{ij} + \frac{\lambda\kappa^2}{2\mu\sqrt{g}} g^{ij}
\pi_B \,.
\ee
The case $\mu=0$ will be discussed later in Sec.~\ref{sec:2.5}.

The Poisson brackets are postulated in the form
(equal time $t$ is understood)
\bea
&& \{ g_{ij}(\vect{x}), \pi^{kl}(\vect{y}) \} = \delta_{(i}^{(k}
\delta_{j)}^{l)} \delta(\vect{x} - \vect{y})\, ,\nn
&& \{ N(\vect{x}), \pi_N(\vect{y}) \} = \delta(\vect{x} - \vect{y})\,
,\quad 
\{ N^i(\vect{x}), \pi_j(\vect{y}) \} = \delta^i_j \delta(\vect{x} -
\vect{y})\, ,\nn
&& \{ A(\vect{x}), \pi_A(\vect{y}) \} = \delta(\vect{x} - \vect{y})\,
,\quad 
\{ B(\vect{x}), \pi_B(\vect{y}) \} = \delta(\vect{x} - \vect{y})\, .
\eea
All the other Poisson brackets between the variables vanish. 
We shall continue to omit the argument $(\vect{x})$ of the fields 
when there is no risk of confusion.

In the following analysis we may lower and raise spatial indices
($i,j,\ldots$) 
with the spatial metric $g_{ij}$ and its inverse $g^{ij}$, e.g. 
$\pi_{ij} = g_{ik} g_{jl} \pi^{kl}$, and we will denote
\be
\pi = g_{ij} \pi^{ij} \, .
\ee
For some integrals over the space $\Sigma_t$, 
we will routinely perform integration by parts, then apply the
divergence 
form of the Stokes theorem and assume that the resulting boundary 
integrals over $\p\Sigma_t$ vanish so that they can be ignored. This 
can be justified by assuming appropriate boundary conditions for the 
variables (asymptotic behaviour at infinity), similarly as in GR.

In order to obtain the Hamiltonian, we first
solve \eqref{pi^ij}--\eqref{pi_B} for $K_{ij}$ and $\dot{B}$,
\bea
K_{ij} &=& \frac{\kappa^2}{\sqrt{g}} \left[ \frac{1}{B} \left( \pi_{ij}
- \frac{1}{3} g_{ij} \pi \right) -
\frac{1}{6\mu} g_{ij} \pi_B \right] \, ,\\
\dot{B} &=& N^i \p_i B - N \frac{\kappa^2}{\sqrt{g}} \left(
\frac{1}{3\mu}\pi +
\frac{1-3\lambda}{6\mu^2} B \pi_B \right) \, ,\label{dotB}
\eea 
and further obtain $\dot{g}_{ij} = 2NK_{ij} + 2\nabla_{(i} N_{j)}$.
Therefore both $g_{ij}$ and $B$ are dynamical
variables and no more primary constraints are needed. The
Hamiltonian is then defined
\be \label{Ha}
H = \int \rd^3 \vect{x} \left( \pi^{ij} \dot{g}_{ij} + \pi_B \dot{B}
\right) - L 
= \int \rd^3 \vect{x} \left( N \mc{H}_0 + N^i \mc{H}_i \right)\, ,
\ee
where the Lagrangian $L$ is given by the action
\eqref{HLFR_action_final},
$S_F = \int\rd t L$, and the so-called Hamiltonian
constraint and the momentum constraints are found to be
\bea
\mc{H}_0 &=& \frac{\kappa^2}{\sqrt{g}} \left[ \frac{1}{B} 
\left( \pi_{ij} \pi^{ij} - \frac{1}{3}\pi^2 \right)
- \frac{1}{3\mu} \pi \pi_B
- \frac{1-3\lambda}{12\mu^2} B \pi_B^2 \right] \nn
&& + \frac{\sqrt{g}}{\kappa^2} \left[ B \left( \cLR + A \right)
- F(A) + 2\mu \Delta B \right] \, ,\nn
\mc{H}_i &=& - 2g_{ij}\nabla_k \pi^{jk} + \nabla_i B \pi_B \nn
&=& -2g_{ij}\p_k \pi^{jk} - \left( 2\p_j g_{ik} - \p_i
g_{jk} \right) \pi^{jk} + \p_i B \pi_B \, ,\label{Hb}
\eea
respectively. We define the total Hamiltonian by
\be \label{H_T}
H_T = H + \int \rd^3 \vect{x} \left( \lambda_N \pi_N + \lambda^i \pi_i
+ \lambda_A \pi_A \right)\, ,
\ee
where the primary constraints \eqref{p_constraints} are multiplied
by the Lagrange multipliers $\lambda_N$, $\lambda^i$, $\lambda_A$. The
total Hamiltonian \eqref{H_T} generates the time evolution of dynamical
variables:
\be
\dot{f}(\vect{x}) = \{ f(\vect{x}), H_T \} \,.
\ee

The primary constraints \eqref{p_constraints} have to be preserved
under the time evolution of the system:
\bea
\dot{\pi}_N &=& \{ \pi_N, H_T \} = - \mc{H}_0 \, ,\nn
\dot{\pi}_i &=& \{ \pi_i, H_T \} = - \mc{H}_i \, ,\nn
\dot{\pi}_A &=& \{ \pi_A, H_T \} = \frac{\sqrt{g}}{\kappa^2} N
\left( - B + F'(A) \right)\, .
\eea
Therefore we impose the secondary constraints:
\bea
&& \mc{H}_0(\vect{x}) \approx 0 \, ,\qquad \mc{H}_i(\vect{x}) \approx 0
\, ,\nn
&& \Phi_A(\vect{x}) \equiv B(\vect{x}) - F'(A(\vect{x}))
\approx 0 \, .\label{s_constraints}
\eea
The Hamiltonian constraint $\mc{H}_0(\vect{x})$, the momentum
constraints 
$\mc{H}_i(\vect{x})$ and the constraint $\Phi_A(\vect{x})$, are all
local. 
It is convenient to introduce globalized versions of the Hamiltonian
and 
momentum constraints:
\bea \label{s_constraints_globalized}
\Phi_0(\eta) &\equiv& \int \rd^3 \vect{x}\eta \mc{H}_0 \approx 0 \, ,\nn
\Phi_S(\xi^i) &\equiv& \int \rd^3 \vect{x}\xi^i \mc{H}_i \approx 0 \, ,
\eea
where $\eta$ and $\xi^i, i=1,2,3$ are arbitrary smearing functions that 
vanish rapidly enough at infinity --- the choices 
$\eta = \delta(\vect{x} - \vect{y})$ and 
$\xi^i = \delta^i_j \delta(\vect{x}-\vect{y})$ will produce the local 
constraints $\mc{H}_0$ and $\mc{H}^j$, which in turn imply the 
smeared constraints.

\subsection{Consistency of the secondary constraints under dynamics}
\label{sec:2.3}

The total Hamiltonian \eqref{H_T} can be rewritten in terms of the 
Hamiltonian and momentum constraints \eqref{s_constraints_globalized} as
\be \label{H_T_as_constraints}
H_T = \Phi_0(N) + \Phi_S(N^i) + \int \rd^3 \vect{x} 
\left( \lambda_N \pi_N + \lambda^i \pi_i
+ \lambda_A \pi_A \right)\, .
\ee
The consistency of the system requires that also the secondary
constraints $\Phi_0(\eta)$, $\Phi_S(\xi^i)$ and $\Phi_A(\vect{x})$ have
 to be preserved under time evolution generated by the total Hamiltonian
\eqref{H_T_as_constraints}:
\bea \label{s_constraints_consistency}
\dot{\Phi}_0(\eta) &=& \{ \Phi_0(\eta), \Phi_0(N) \} + \{ \Phi_0(\eta),
\Phi_S(N^i) \}
+ \int\rd^3 \vect{x} \lambda_A(\vect{x}) \{ \Phi_0(\eta),
\pi_A(\vect{x}) \} \approx 0\, ,\nn
\dot{\Phi}_S(\xi^i) &=& \{ \Phi_S(\xi^i), \Phi_0(N) \} + \{
\Phi_S(\xi^i), \Phi_S(N^i) \} \approx 0\,,\\
\dot{\Phi}_A(\vect{x}) &=& \{ \Phi_A(\vect{x}), \Phi_0(N) \} + \{
\Phi_A(\vect{x}), \Phi_S(N^i) \} 
+ \int\rd^3 \vect{y} \lambda_A(\vect{y}) \{ \Phi_A(\vect{x}), 
\pi_A(\vect{y}) \} \approx 0 \, ,\nonumber
\eea
where we have used the fact that the constraints $\pi_N$ and 
$\pi_i$ have strongly vanishing Poisson brackets with every constraint, 
and that
\be
\{ \Phi_S(\xi^i), \pi_A \} = 0 \, .
\ee
We need to calculate the rest of the algebra of the constraints under
the
Poisson bracket. The Poisson brackets between the momentum constraint
$\Phi_S(\xi^i)$ and the canonical variables are
\bea \label{diffeom_generator}
\{ \Phi_S(\xi^i), B \} &=& - \xi^i \p_i B\, ,\nn \{ \Phi_S(\xi^i),
\pi_B \} &=& - \p_i \left( \xi^i \pi_B \right)\, ,\nn
\{ \Phi_S(\xi^k), g_{ij} \} &=& - \xi^k \p_k g_{ij} - g_{ik} \p_j
\xi^k - g_{jk} \p_i \xi^k\, ,\nn
\{ \Phi_S(\xi^k), \pi^{ij} \} &=& - \p_k \left( \xi^k \pi^{ij} \right) +
\pi^{ik} \p_k \xi^j +\pi^{jk}
\p_k \xi^i\, ,
\eea
and trivially zero for $A$ and $\pi_A$,
\be \label{diffeom_generator2}
\{ \Phi_S(\xi^i), A \} = 0\, ,\quad \{ \Phi_S(\xi^i), \pi_A \} = 0 \, .
\ee
Thus $\Phi_S(\xi^i)$ generates the spatial diffeomorphisms for the 
variables $B, \pi_B, g_{ij}, \pi^{ij}$, and consequently for any
function 
or functional constructed from these variables, and treats the
variables 
$A,\pi_A$ as constants. By using this result
\eqref{diffeom_generator}--\eqref{diffeom_generator2} we obtain 
the Lie algebra of the generators $\Phi_S(\xi^i)$:
\be
\{ \Phi_S(\xi^i), \Phi_S(\eta^i) \} = \Phi_S(\xi^j \p_j
\eta^i - \eta^j \p_j \xi^i) \approx 0 \, .\label{PhiS_PBs}
\ee
Then we calculate their Poisson brackets with the Hamiltonian constraint
$\Phi_0(\eta)$:
\be
\{ \Phi_S(\xi^i), \Phi_0(\eta) \} = \Phi_0(\xi^i \p_i \eta) \approx 0 \,
,
\ee
which tells us that $\mc{H}_0$ is a scalar density under spatial
diffeomorphism. 
The momentum constraints $\mc{H}_i(\vect{x})$ will be first-class
everywhere 
and consistent under time evolution.

Then consider the constraint $\Phi_A(\vect{x})$, whose nonvanishing 
Poisson brackets are
\bea
\{ \Phi_A(\vect{x}), \Phi_0(\eta) \} &=& - \eta
\frac{\kappa^2}{\sqrt{g}} 
\left( \frac{1}{3\mu}\pi + \frac{1-3\lambda}{6\mu^2}B\pi_B \right) \,
,\nn
\{ \Phi_A(\vect{x}), \Phi_S(\xi^i) \} &=& \xi^i \p_i B \, ,\nn
\{ \Phi_A(\vect{x}), \pi_A(\vect{y}) \} &=& - F''(A(\vect{x}))
\delta(\vect{x}-\vect{y}) \, . \label{PhiA_PBs}
\eea
Thus, in order to satisfy the consistency conditions
 \eqref{s_constraints_consistency}, we have to impose the tertiary
constraint
\be \label{lambdaA_constraint}
N^i \p_i B - N \frac{\kappa^2}{\sqrt{g}} \left( \frac{1}{3\mu}\pi +
\frac{1-3\lambda}{6\mu^2}B\pi_B \right)
 - \lambda_A F''(A) \approx 0\, .
\ee
Since $F''(A)=0$ would essentially reproduce the usual non-projectable
\HL gravity, we assume that $F''(A) \neq 0$.
The first two terms in \eqref{lambdaA_constraint}, i.e. the expression
for $\dot{B}$ in \eqref{dotB}, does not vanish due to the
established constraints \eqref{p_constraints} and \eqref{s_constraints}.
Therefore \eqref{lambdaA_constraint} is a restriction on the Lagrange
multiplier $\lambda_A$, and we can solve it:
\be \label{lambda_A}
\lambda_A = \frac{1}{F''(A)} \left( N^i \p_i B - N
\frac{\kappa^2}{\sqrt{g}} \left( \frac{1}{3\mu}\pi +
\frac{1-3\lambda}{6\mu^2}B\pi_B \right) \right)\, .
\ee
Introducing \eqref{lambda_A} into the Hamiltonian
\eqref{H_T_as_constraints} ensures that the constraint
$\Phi_A(\vect{x})$
is consistent.

Finally we consider the Hamiltonian constraint $\Phi_0(\eta)$. The
Poisson 
bracket with the primary constraint $\pi_A(\vect{x})$ vanishes in
\eqref{s_constraints_consistency}
\be \label{PB_Phi0_piA}
\{ \Phi_0(\eta), \pi_A(\vect{x}) \} = \eta \frac{\sqrt{g}}{\kappa^2}
\Phi_A(\vect{x}) \approx 0 \, .
\ee
Also the Poisson bracket of the Hamiltonian constraint $\Phi_0(\eta)$ 
with itself has to be calculated in order to check its consistency.
Since the 
Poisson bracket is antisymmetric with respect to its arguments and the
Hamiltonian constraint does not depend on the primary constraints
\eqref{p_constraints}, 
we can write the Poisson bracket of Hamiltonian constraints as
\be
\{ \Phi_0(\xi), \Phi_0(\eta) \} = \int\rd^3\vect{x} \left( 
\frac{\delta \Phi_0(\xi)}{\delta g_{ij}(\vect{x})} 
\frac{\delta \Phi_0(\eta)}{\delta \pi^{ij}(\vect{x})} 
+ \frac{\delta \Phi_0(\xi)}{\delta B(\vect{x})} 
\frac{\delta \Phi_0(\eta)}{\delta \pi_B(\vect{x})} \right) 
- (\xi \leftrightarrow \eta)
\ee
or
\bea \label{PB_of_Phi0s}
\{ \Phi_0(\xi), \Phi_0(\eta) \} &=& \int\rd^3\vect{x} \left( 
\{ \Phi_0(\xi), \pi^{ij}(\vect{x}) \} \{ g_{ij}(\vect{x}), \Phi_0(\eta)
\} \right.\nn
&&\qquad\quad + \left. \{ \Phi_0(\xi), \pi_B(\vect{x}) \} \{
B(\vect{x}), \Phi_0(\eta) \} \right)
- (\xi \leftrightarrow \eta) \, .
\eea
Further simplification follows from the fact that the Poisson brackets
of the Hamiltonian constraint $\Phi_0(\eta)$ with the field variables
$g_{ij}$ and $B$ are proportional to $\eta$, i.e. of the form $\eta f$
where $f$ is a function of the canonical variables:
\bea \label{PB_g_ij_Phi0}
\{ g_{ij}(\vect{x}), \Phi_0(\eta) \} &=& \eta \frac{\kappa^2}{\sqrt{g}}
\left[ \frac{2}{B} \left( \pi_{ij} - \frac{1}{3} g_{ij} \pi \right)
- \frac{1}{3\mu} g_{ij} \pi_B \right] = 2\eta K_{ij} \, ,\\
\{ B(\vect{x}), \Phi_0(\eta) \} &=& - \eta \frac{\kappa^2}{\sqrt{g}}
\left( \frac{1}{3\mu} \pi + \frac{1-3\lambda}{6\mu^2} B\pi_B \right) \,
.
\eea
Therefore, the parts of the Poisson brackets of the Hamiltonian
constraint $\Phi_0(\xi)$ with the momenta $\pi^{ij}$ and $\pi_B$ that
are proportional to $\xi$ will contribute to terms that are proportional
to $\xi\eta=\eta\xi$. These terms will necessarily cancel out due to the
antisymmetry of \eqref{PB_of_Phi0s} under $\xi \leftrightarrow \eta$. 
Thus, in order to calculate \eqref{PB_of_Phi0s}, we only need those
parts of the Poisson brackets with the momenta that contain spatial
derivatives of $\xi$. 
These Poisson brackets are found to be
\bea \label{PB_Phi0_pi^ij}
\{ \Phi_0(\xi), \pi^{ij}(\vect{x}) \} &=& - \xi
\frac{\kappa^2}{\sqrt{g}} 
g^{ij} \left[ \frac{1}{2B} \left( \pi_{kl} \pi^{kl}
 - \frac{1}{3}\pi^2 \right) \right. - \left. \frac{1}{6\mu} \pi \pi_B -
\frac{1-3\lambda}{24\mu^2} B \pi_B^2 \right] \nn
&+& \xi \frac{\sqrt{g}}{\kappa^2} g^{ij} \left[ \frac{1}{2} B \left(
\cLR[g_{kl}] + A \right)
 - \frac{1}{2} F(A) + \mu \Delta B \right] \nn
&+& \xi \frac{\kappa^2}{\sqrt{g}} \left[ \frac{2}{B}\left(
\pi^i_{\phantom{i}k} \pi^{jk} - \frac{1}{3} \pi^{ij} \pi \right) -
\frac{1}{3\mu} \pi^{ij}\pi_B \right] \\
&+& \frac{1}{\kappa^2} \int\rd^3\vect{y}\sqrt{g} \xi(\vect{y}) \left(
B(\vect{y}) \frac{\delta \cLR[g_{kl}(\vect{y})]}{\delta
g_{ij}(\vect{x})} + 2\mu \frac{\delta \Delta B(\vect{y})}{\delta
g_{ij}(\vect{x})} \right) \nonumber
\eea
and
\bea \label{PB_Phi0_piB}
\{ \Phi_0(\xi), \pi_B(\vect{x}) \} &=& - \xi \frac{\kappa^2}{\sqrt{g}}
\left[ \frac{1}{B^2} 
\left( \pi_{ij} \pi^{ij} - \frac{1}{3}\pi^2 \right) +
\frac{1-3\lambda}{12\mu^2} \pi_B^2 \right] \nn
&+& \xi \frac{\sqrt{g}}{\kappa^2} \left( \cLR + A \right) +
\frac{\sqrt{g}}{\kappa^2} 2\mu \Delta \xi \, ,
\eea
where the last term was obtained from
\be
\{ \int\rd^3\vect{y} \xi \frac{\sqrt{g}}{\kappa^2} 2\mu \Delta B,
\pi_B(\vect{x}) \} = \frac{2\mu}{\kappa^2} \int\rd^3\vect{y}\sqrt{g}
\xi(\vect{y}) \frac{\delta \Delta B(\vect{y})}{\delta B(\vect{x})} =
\frac{\sqrt{g}}{\kappa^2} 2\mu \Delta \xi(\vect{x}) \, .
\ee
We find immediately that only the last term in both
\eqref{PB_Phi0_pi^ij} and 
\eqref{PB_Phi0_piB} give nonzero contributions to the Poisson 
bracket \eqref{PB_of_Phi0s}.

Then we must calculate the variations of the potential part $\cLR$ 
and $\Delta B$ with respect to the spatial metric in
\eqref{PB_Phi0_pi^ij}. For this we need the variations of the involved
geometrtic quanties. The variations of the connection coefficients
$\Gamma_{ij}^k$, the Ricci tensor $R_{ij}$ and the scalar curvature $R$
are given by
\bea
\delta \Gamma_{ij}^k &=& \frac{1}{2} g^{kl} \left( \nabla_i \delta
g_{lj} + 
\nabla_j \delta g_{il} - \nabla_l \delta g_{ij} \right) ,\nn
\delta R_{ij} &=& \frac{1}{2} g^{kl} \left( \nabla_k \nabla_i \delta
g_{lj} + \nabla_k \nabla_j \delta g_{li} - \nabla_i \nabla_j \delta
g_{kl} \right) - \frac{1}{2} \Delta \delta g_{ij} ,\nn
\delta R &=& - R^{ij} \delta g_{ij} + \nabla^i \nabla^j \delta g_{ij} -
g^{ij} \Delta \delta g_{ij} .
\eea
For $\Delta B$ we obtain the variation
\be
\delta\Delta B = - \delta g_{ij} \nabla^i \nabla^j B + \frac{1}{2} 
g^{ij} \left( \nabla_i \delta g_{kj} + \nabla_j \delta g_{ik} - \nabla_k
\delta g_{ij} \right) \nabla^k B
\ee
and then in (\ref{PB_Phi0_pi^ij}) we have
\bea
a_{ij}(\vect{x}) \int\rd^3\vect{y} \sqrt{g} \xi(\vect{y}) 2\mu
\frac{\delta\Delta B(\vect{y})}{\delta g_{ij}(\vect{x})}
&=& a_{ij} 2\mu \sqrt{g} \left[ - 2\xi \nabla^{(i} \nabla^{j)} B -
\nabla^{(i} \xi \nabla^{j)} B \right.\nn
&+& \left. \frac{1}{2} g^{ij} \left( \nabla_k \xi \nabla^k B + \xi
\Delta B \right) \right] ,
\eea
where $a_{ij}$ represents any tensor, and the first term and the last
term will be cancelled for the familiar reason. 
Thus we obtain the general result for the Poisson bracket of Hamiltonian
constraints \eqref{PB_of_Phi0s}:
\bea \label{PB_of_Phi0s_2nd}
\{ \Phi_0(\xi), \Phi_0(\eta) \} &=& \int\rd^3\vect{x} \eta \left\{ 
\left[ \frac{2}{B} \left( \pi_{ij} - \frac{1}{3} g_{ij} \pi \right) 
- \frac{1}{3\mu} g_{ij} \pi_B \right]  \left( \mu g^{ij} \nabla^k B
\nabla_k \xi \right.\right.\nn
&-& \left. 2\mu \nabla^{(i} B \nabla^{j)} \xi + \frac{1}{\sqrt{g}}
\int\rd^3\vect{y}\sqrt{g} \xi(\vect{y}) B(\vect{y}) \frac{\delta
\cLR[g_{kl}(\vect{y})]}{\delta g_{ij}(\vect{x})} \right) \nn
&-& \left. \left( \frac{2}{3} \pi + \frac{1-3\lambda}{3\mu} B\pi_B
\right) 
\Delta \xi \right\} - (\xi \leftrightarrow \eta) \, .
\eea
Further progress can be obtained by specifying the form of the potential
part $\cLR$ in the action.

For $\cLR$ let us start with the simplest case that could be the
low-energy effective potential
\be \label{cLR_1st}
\cLR = \alpha_0 + \alpha_1 R  \, ,
\ee
where $\alpha_0$ and $\alpha_1$ are coupling constants. \eqref{cLR_1st}
gives 
$\tildeR$ in \eqref{tildeR_2nd} as a ``deformed scalar curvature +
constant'':
\be
\tildeR = K^{ij} K_{ij} - \lambda K^2 - \alpha_1 R + 2\mu \fnabla_\mu
\left(
n^\mu K \right) - \frac{2\mu}{N} \Delta N - \alpha_0 \, .
\ee
When $\lambda=1$, $\mu=1$, $\alpha_1=-1$ and $\alpha_0=0$, this
$\tildeR$ reduces to the scalar curvature $\fR$ of space-time. For
asymptotically flat spaces we assume $\alpha_0=0$. For \eqref{cLR_1st}
we obtain
\be
a_{ij}(\vect{x}) \int\rd^3\vect{y} \sqrt{g} \xi(\vect{y}) B(\vect{y})
\frac{\delta \cLR[g_{kl}(\vect{y})]}{\delta g_{ij}(\vect{x})}
= a_{ij} \alpha_1 \sqrt{g} \left[ - \xi B R^{ij} + \nabla^{(i}
\nabla^{j)} (\xi B) - g^{ij} \Delta (\xi B) \right] ,
\ee
where $a_{ij}$ represents any tensor, and again the parts that contain
no derivatives of $\xi$ will cancel out in the Poisson bracket
\eqref{PB_of_Phi0s_2nd}.
This expression contains first and second order spatial derivatives of
$\xi$. 
Thus we obtain the result
\be \label{PB_of_Phi0s_3rd}
\{ \Phi_0(\xi), \Phi_0(\eta) \} = \int\rd^3\vect{x} \eta \left( 
C_2^{ij} \nabla_{(i}\nabla_{j)} \xi + C_1^i \nabla_i \xi
\right) - (\xi \leftrightarrow \eta) \, ,
\ee
where we have defined:
\bea \label{C_2_1}
C_2^{ij} &=& 2\alpha_1 \pi^{ij} - g^{ij} \left( \frac{2(\alpha_1+1)}{3}
\pi + \frac{1-2\alpha_1-3\lambda}{3\mu} B\pi_B \right) \, ,\nn
C_1^i &=& \left[ 4(\alpha_1-\mu) \pi^{ij} + g^{ij} \left(
\frac{4(-\alpha_1+\mu)}{3} \pi + \frac{4\alpha_1-\mu}{3\mu} B \pi_B
\right) \right] \frac{\nabla_j B}{B} \, .
\eea
The tensor $C_2^{ij}$ is symmetric and so is the part of $C_1^i$ that
multiplies $\nabla_j B$ in the definition \eqref{C_2_1}.
Note that in the case $\alpha_1=-1, \lambda=1$ the coefficient of the
$\Delta\xi$-term vanishes in \eqref{PB_of_Phi0s_3rd} so that $C_2^{ij}
\rightarrow 2\alpha_1 \pi^{ij}$. Finally we can write
\eqref{PB_of_Phi0s_3rd} into a form where the integrand is proportional
to $\xi$:
\be \label{PB_of_Phi0s_4th}
\{ \Phi_0(\xi), \Phi_0(\eta) \} = \int\rd^3\vect{x} \xi \left( E_1^i
\nabla_i \eta + E_0 \eta \right) \, .
\ee
where we have defined
\bea \label{E_1_0}
E_1^i &=& 2 \left( \nabla_j C_2^{ij} - C_1^i \right) \,, \nn
E_0 &=& \nabla_{(i} \nabla_{j)} C_2^{ij} - \nabla_i C_1^i = \frac{1}{2}
\nabla_i E_1^i \,.
\eea
Note that the terms involving the second spatial derivative $\nabla_{(i}
\nabla_{j)} \eta$ cancel each other independently of the form of
$C_2^{ij}$. 
Now the condition that is necessary for the preservation of the local
Hamiltonian constraint $\mc{H}_0(\vect{x})$ can be obtained by inserting
$\xi = \delta(\vect{x}-\vect{y})$ and $\eta = N$ into the Poisson
bracket \eqref{PB_of_Phi0s_4th}. 
Thus, in order to ensure the consistency of the Hamiltonian constraint 
$\mc{H}_0(\vect{x})$, we must impose the constraint
\be \label{constraint_on_N}
\tilde{E}_1^i \nabla_i N + \tilde{E}_0 N \approx 0 \, ,
\ee
where we have factored out some parts of the $E_n$'s that vanish due to
the momentum constraints by defining:
 \bea \label{tildeE_1_0}
\tilde{E}_1^i &=& E_1^i + 2 \alpha_1 \mc{H}^i \nn
&=& 2\left[ - \frac{2(\alpha_1+1)}{3}\nabla^i \pi 
+ 4(-\alpha_1+\mu) \left( \pi^{ij} - \frac{1}{3}g^{ij} \pi \right)
\frac{\nabla_j B}{B}  \right. \nn
&-& \left. \frac{1-2\alpha_1-3\lambda}{3\mu} B \nabla^i \pi_B + \left(
\frac{3\lambda+\mu-2\alpha_1-1}{3\mu}+\alpha_1 \right) \nabla^i B \pi_B
\right] \, ,\nn
\tilde{E}_0 &=& E_0 + \alpha_1 \nabla_i \mc{H}^i = \frac{1}{2} \nabla_i
\tilde{E}_1^i \, .
\eea
This is a homogeneous first-order partial differential equation for the
lapse $N$. 
As such it always has the solution $N=0$. 
Due to the relation \eqref{E_1_0} of $E_0$ and $E_1^i$, we can rewrite
the constraint \eqref{constraint_on_N} as a divergence (after
multiplying it by $2N$)
\be \label{constraint_on_N_2nd}
\nabla_i \left( N^2 \tilde{E}_1^i \right) \approx 0 \, .
\ee

Let us next consider the case with the critical exponent $z=3$ that
could provide an UV complete theory. For the potential $\cLR$ there are
many terms that have the same scaling dimension as the kinetic terms
under the anisotropic scaling \eqref{anisotropic_scaling} with $z=3$. 
Such terms are, for example, the terms quadratic in curvature
\be \label{3rd_deriv_delta}
\nabla^i R \nabla_i R \, ,\quad \nabla^i R^{jk} \nabla_i R_{jk} \,
,\quad \nabla^i R^{jk} \nabla_j R_{ki}
\ee
and
\be
R \Delta R \, ,\quad R^{ij} \Delta R_{ij} \, ,\quad R^{ij} \nabla_i
\nabla_j R \, ,\label{4th_deriv_delta}
\ee
which modify the propagator in addition to providing interactions, and
the terms cubic in curvature that are pure interactions
\be
R^3 \, ,\quad R R^{ij} R_{ij} \, ,\quad R^i_{\phantom{i}j}
R^j_{\phantom{j}k} R^k_{\phantom{k}i} \, .\label{cubic_interaction}
\ee
Many of the terms are related to each other due to the properties of the
Riemann tensor in three dimensions (Weyl tensor vanishes), the Bianchi
identity, and integration by parts. Indeed only two terms of the types
\eqref{3rd_deriv_delta} and \eqref{4th_deriv_delta} are independent,
which we choose to be
\be
\nabla^i R^{jk} \nabla_i R_{jk} \, ,\quad R \Delta R \,.
\ee
The most general potential that contains all the independent
renormalizable and super-renormalizable terms, while maintaining the
canonical form of the Hamiltonian \eqref{Ha}, is
\bea \label{cLR_general}
\cLR &=& \alpha_0 + \alpha_1 R +\alpha_2 R^2 + \alpha_3 R^{ij} R_{ij} +
\alpha_4 R^3 + \alpha_5 R R^{ij} R_{ij} \nn
&+& \alpha_6 R^i_{\phantom{i}j} R^j_{\phantom{j}k} R^k_{\phantom{k}i} +
\alpha_7 \nabla^i R^{jk} \nabla_i R_{jk} + \alpha_8 R \Delta R \,.
\eea
Under the variation of the spatial metric the terms involving second
order spatial derivatives of curvature \eqref{4th_deriv_delta} contain
spatial derivatives of the variation $\delta g_{ij}$ up to fourth order.
Likewise, the variations of the terms \eqref{3rd_deriv_delta} contain
spatial derivatives of the variation $\delta g_{ij}$ up to third order,
and the variation of \eqref{cubic_interaction} up to second order. This
means that the variation of the potential \eqref{cLR_general} contains
spatial derivatives of the variation $\delta g_{ij}$ up to fourth order.
Thus the Poisson bracket of Hamiltonian constraints has the form
\be \label{PB_of_Phi0s_5th}
\{ \Phi_0(\xi), \Phi_0(\eta) \} = \int\rd^3\vect{x} \eta \left( 
C_4^{ijkl} \nabla_{ijkl} \xi + C_3^{ijk} \nabla_{ijk} \xi + C_2^{ij}
\nabla_{ij} \xi + C_1^i \nabla_i \xi
\right) - (\xi \leftrightarrow \eta) \, ,
\ee
where $C_n$'s are symmetric tensors consisting of the canonical
variables and their spatial derivatives, and we denote
$\nabla_{ij}=\nabla_{(i}\nabla_{j)}$ etc. For arbitrary couplings
$\alpha_m$ the tensors $C_n$ have quite complicated forms, which we do
not present here. After integration by parts we obtain
\bea \label{PB_of_Phi0s_6th}
\{ \Phi_0(\xi), \Phi_0(\eta) \} = \int\rd^3\vect{x} \xi \left( E_3^{ijk}
\nabla_{ijk} \eta + E_2^{ij} \nabla_{ij} \eta + E_1^i \nabla_i \eta +
E_0 \eta \right) \, ,
\eea
where we have defined
\bea \label{E_n}
E_3^{ijk} &=& 4\nabla_l C_4^{ijkl} - 2 C_3^{ijk}\,,\nn
E_2^{ij} &=& 6\nabla_{kl} C_4^{ijkl} - 3\nabla_k C_3^{ijk} = \frac{3}{2}
\nabla_k E_3^{ijk}\,,\nn
E_1^i &=& 4\nabla_{jkl} C_4^{ijkl} - 3\nabla_{jk} C_3^{ijk} + 2\nabla_j
C_2^{ij} - 2 C_1^i \,,\nn
E_0 &=& \nabla_{ijkl} C_4^{ijkl} - \nabla_{ijk} C_3^{ijk} + \nabla_{ij}
C_2^{ij} - \nabla_i C_1^i \,.
\eea
The resulting tertiary constraint is of the form
\be \label{constraint_on_N_generic}
E_3^{ijk} \nabla_{ijk} N + E_2^{ij} \nabla_{ij} N + E_1^i \nabla_i N+
E_0 N \approx 0 \, .
\ee
The condition \eqref{constraint_on_N_generic} is again homogeneous in
$N$ and is therefore satisfied by $N=0$. 
Note that the fourth order spatial derivative of $N$ cancels out
similarly as the second order derivative did in the case
\eqref{cLR_1st}. More generally for every even $n$, $C_n$ contributes to
$E_k$ for $k<n$, but never to $E_n$. For every uneven $n$, the
contribution of $C_n$ to $E_n$ is $-2C_n$. Thus the highest spatial
derivative of $N$ in the constraints like \eqref{constraint_on_N} and
\eqref{constraint_on_N_generic} is always of uneven order. This is
clearly a very general result that holds in any theory where the Poisson
bracket of Hamiltonian constraints is of the form
\eqref{PB_of_Phi0s_5th} with nonvanishing coefficients $C_n$ usually up
to $n=z+1$ ---  including the usual \HL gravity. 
The constraints \eqref{constraint_on_N_2nd} and
\eqref{constraint_on_N_generic} will be discussed further in
Sec.~\ref{sec:2.7}.

\subsection{Elimination of the auxiliary field \texorpdfstring{$A$}{A}}
\label{sec:2.4}

According to the Poisson brackets between the constraints 
\eqref{PhiS_PBs}--\eqref{PhiA_PBs} and \eqref{PB_Phi0_piA}, we can set 
the second-class constraints $\pi_A(\vect{x})$ and $\Phi_A(\vect{x})$ 
to vanish strongly. Then the Hamiltonian constraint $\Phi_0(\eta)$ 
will have weakly vanishing Poisson brackets with every constraint
except 
with itself \eqref{PB_of_Phi0s_6th}. The momentum constraints 
$\Phi_S(\xi^i)$ will be first-class assuming its Poisson bracket with 
possible additional constraints are weakly vanishing too. To this end, 
we replace the Poisson bracket with the Dirac bracket, which is given by
\bea \label{DB}
\{ f(\vect{x}), h(\vect{y}) \}_\mr{DB} = \{ f(\vect{x}),
h(\vect{y}) \} &+& \int\rd^3 \vect{z} \frac{1}{F''(A(\vect{z}))}
\left( \{ f(\vect{x}), \pi_A(\vect{z}) \} \{ \Phi_A(\vect{z}),
h(\vect{y}) \} \right.\nn
&&\qquad\qquad - \left. \{ f(\vect{x}), \Phi_A(\vect{z}) \} \{
\pi_A(\vect{z}), h(\vect{y}) \} \right) \, ,
\eea
where $f$ and $h$ are any functions of the canonical variables. 
Assuming we can solve the constraint $\Phi_A(\vect{x})=0$, i.e. 
$B=F'(A)$, for $A=\tilde{A}(B)$, where $\tilde{A}$ is the inverse of 
the function $F'$, we can eliminate the variables $A$ and $\pi_A$. 
Thus the final variables of the system are $g_{ij}, \pi^{ij}, B,
\pi_B$. 
The lapse $N$ and the shift vector $N^i$, together with $\lambda_N$ 
and $\lambda^i$, are non-dynamical multipliers. Since every dynamical 
variable has a vanishing Poisson bracket with the constraint $\pi_A$, 
the Dirac bracket \eqref{DB} reduces to the Poisson bracket,
\be \label{DB_equal_PB}
\{ f(\vect{x}), h(\vect{y}) \}_\mr{DB} = \{ f(\vect{x}), h(\vect{y})
\}\, .
\ee

\subsection{The cases \texorpdfstring{$F''=0$}{F''=0} and
\texorpdfstring{$\mu=0$}{mu=0}}
\label{sec:2.5}

For $F''=0$, i.e. $F(\tildeR) = c_1 \tildeR + c_0$ with constants $c_1$
and $c_0$, the action \eqref{HLFR_action} with \eqref{cLR_general}
reduces to the action of the usual \HL gravity without the detailed
balance condition, when the couplings are redefined: $c_1/\kappa^2
\rightarrow 1/\kappa^2$, $(c_1\alpha_0 - c_0)/\kappa^2 \rightarrow
\alpha_0$ and $c_1\alpha_n/\kappa^2 \rightarrow \alpha_n$ for
$n=1,2,\ldots,8$. 
The value of the parameter $\mu$ is irrelevant, since the term involving
(\ref{mu-term}) is a total divergence and hence it can be dropped. For
Hamiltonian analysis of non-projectable \HL gravity see
Ref.~\cite{Henneaux:2010} and references therein, recalling the
observations about the general structure of the Poisson bracket of
Hamiltonian constraints made in Sec.~\ref{sec:2.3}.

Let us then consider the case $\mu=0$, and assume $F'' \neq 0$. As we
mentioned this is a generalization of the (second) model proposed in
Ref.~\cite{Kluson:2009xx} without the detailed balance condition. The
additional primary constraint $\pi_B(\vect{x}) \approx 0$ has to be
introduced in addition to \eqref{p_constraints}. The canonical momenta
are
\be
\pi^{ij} = \frac{1}{\kappa^2} \sqrt{g}B \cG^{ijkl} K_{kl} \,.
\ee
The field $B$ is non-dynamical. For $\lambda \neq 1/3$, we obtain the
Hamiltonian and the momentum constraints
\bea \label{mu0_H_0}
\mc{H}_0 &=& \frac{\kappa^2}{\sqrt{g} B} \pi^{ij}\cG_{ijkl}\pi^{kl} +
\frac{\sqrt{g}}{\kappa^2} \left[ B \left( \cLR + A \right) - F(A)
\right] \,,\nn
\mc{H}_i &=& -2g_{ij}\nabla_k \pi^{jk} \,.
\eea
The preservation of the constraint $\pi_B(\vect{x})$ under time
evolution implies the secondary constraint
\be
\Phi_B(\vect{x}) \equiv \frac{\kappa^2}{\sqrt{g} B^2}
\pi^{ij}\cG_{ijkl}\pi^{kl} - \frac{\sqrt{g}}{\kappa^2} \left( \cLR + A
\right) \approx 0 \,.
\ee
The constraints $\chi_a(\vect{x}) = (\pi_A, \Phi_A, \pi_B, \Phi_B)$ are
second-class and the corresponding Dirac bracket is defined
\be \label{mu0_DB}
\{ f(\vect{x}), h(\vect{y}) \}_\mr{DB} = \{
f(\vect{x}), h(\vect{y}) \} - \int \rd^3 \vect{z} \rd^3 \vect{z'} \{
f(\vect{x}), \chi_a(\vect{z}) \} C^{ab}(\vect{z}, \vect{z'})  \{
\chi_b(\vect{z'}), h(\vect{y}) \}
\ee
where $C^{ab}(\vect{z}, \vect{z'}) = C^{ab}(\vect{z})
\delta(\vect{z}-\vect{z'})$ is the inverse of
\be
C_{ab}(\vect{x}, \vect{y}) \equiv \{ \chi_a(\vect{x}), \chi_b(\vect{y})
\} \nn
= C_{ab}(\vect{x}) \delta(\vect{x}-\vect{y}) \,.
\ee
The nonvanishing components of these antisymmetric matrices are
\bea
C_{12}(\vect{x}) = F''(A) \,, && C_{14}(\vect{x}) =
\frac{\sqrt{g}}{\kappa^2} \,,\nn
C_{23}(\vect{x}) = 1 \,, && C_{34}(\vect{x}) =
\frac{2\kappa^2}{\sqrt{g}B^3} \pi^{ij}\cG_{ijkl}\pi^{kl} \,,
\eea
and
\bea
C^{12}(\vect{z}) &=&
-\frac{2\pi^{ij}\cG_{ijkl}\pi^{kl}}{2\pi^{ij}\cG_{ijkl}\pi^{kl} F''(A) +
gB^3/\kappa^4} \,,\nn 
C^{14}(\vect{z}) &=&
-\frac{\sqrt{g}B^3/\kappa^2}{2\pi^{ij}\cG_{ijkl}\pi^{kl} F''(A) +
gB^3/\kappa^4} \,,\nn
C^{23}(\vect{z}) &=& -\frac{gB^3/\kappa^4}{2\pi^{ij}\cG_{ijkl}\pi^{kl}
F''(A) + gB^3/\kappa^4} \,,\nn
C^{34}(\vect{z}) &=&
-\frac{\sqrt{g}F''(A)B^3/\kappa^2}{2\pi^{ij}\cG_{ijkl}\pi^{kl} F''(A) +
gB^3/\kappa^4} \,.
\eea
Since the constraints $\chi_1=\pi_A$ and $\chi_3=\pi_B$ have vanishing
Poisson brackets with the canonical variables $g_{ij}$ and $\pi^{ij}$,
and we have $C^{24}(\vect{x}) =0$, we again find that the Dirac bracket
reduces to the Poisson bracket \eqref{DB_equal_PB} for any functions of
$g_{ij}$ and $\pi^{ij}$. The auxiliary fields can be eliminated from the
action by solving $\Phi_A(\vect{x})=0$ and $\Phi_B(\vect{x})=0$ for
\be
A = \tilde{A}\left( \frac{\kappa^2}{\sqrt{g}}
\pi^{ij}\cG_{ijkl}\pi^{kl}, \frac{\sqrt{g}}{\kappa^2}\cLR \right)
\,,\quad B = F'(\tilde{A}) \,,
\ee
where $\tilde{A}$ satisfies
\be
\frac{\kappa^2}{\sqrt{g} F'(\tilde{A})^2} \pi^{ij}\cG_{ijkl}\pi^{kl} -
\frac{\sqrt{g}}{\kappa^2} \left( \cLR + \tilde{A} \right) = 0 \,.
\ee
Then the Hamiltonian constraint can be written in terms of $g_{ij}$ and
$\pi^{ij}$:
\be
\mc{H}_0 =  \frac{2\kappa^2}{\sqrt{g} F'(\tilde{A})}
\pi^{ij}\cG_{ijkl}\pi^{kl} - \frac{\sqrt{g}}{\kappa^2} F(\tilde{A}) \,.
\ee
The seemingly simple form of this Hamiltonian can be deceiving, since
the form of the function $\tilde{A}$ can be complex. For example, when
$F(R)$ is a polynomial of degree $n$, then $\tilde{A}$ is a root of a
polynomial $p(\tilde{A})$ of degree $2(n-1)+1$.

The momentum constraints \eqref{mu0_H_0} clearly generate the spatial
diffeomorphisms of the canonical variables, since $\mc{H}_i$ are
identical to the momentum constraints of GR. It is also clear that the
Poisson bracket of Hamiltonian constraints has the general form
\eqref{PB_of_Phi0s_5th} when the potential takes the general canonical
form \eqref{cLR_general}. 

Since $\cG^{ijkl} g_{kl} = (1-3\lambda) g^{ij}$, the value $\lambda=1/3$
implies the primary constraint $\pi(\vect{x}) \approx 0$ similarly as in
the usual \HL gravity. Then the Hamiltonian and the momentum constraints
are found to be as in \eqref{mu0_H_0} but with
$\pi^{ij}\cG_{ijkl}\pi^{kl}$ replaced by $\pi_{ij}\pi^{ij}$. The
preservation of $\pi(\vect{x}) \approx 0$ under time evolution is a
nontrivial matter that is not discussed here.

\subsection{On the class of theories with a projectable lapse
\texorpdfstring{$N$}{N}}
\label{sec:2.6}

For comparison we briefly consider the class of theories with a 
projectable lapse $N=N(t)$ (see \cite{Carloni:2010} for more details). 
In this case there is only one (global) Hamiltonian constraint 
$\Phi_0 \equiv \int\rd^3\vect{x} \mc{H}_0 = \Phi_0(1)$. 
Now the Poisson bracket of Hamiltonian constraints vanishes, 
\be \label{Phi0_PB}
\{\Phi_0, \Phi_0\}=0 \,,
\ee
and there is no need for additional constraints like the one 
\eqref{constraint_on_N_generic} in the class of theories with a 
non-projectable lapse.

As explained in Sec.~\ref{sec:2.4}, the second-class constraints 
$\pi_A(\vect{x})$ and $\Phi_A(\vect{x})$ can be set to vanish strongly 
by utilizing the Dirac bracket \eqref{DB}, which reduces to the Poisson 
bracket \eqref{DB_equal_PB}. As a result the Hamiltonian constraint 
$\Phi_0$ and the momentum constraints $\Phi_S(\xi^i)$ are first-class, 
since they have vanishing Dirac brackets with every constraint. 
Finally the total Hamiltonian is the sum of the first-class constraints
\be \label{H_T_sum_of_first-class}
H_T = N\Phi_0 + \Phi_S(N^i) + \lambda_N \pi_N + \int\rd^3 \vect{x}
\lambda^i \pi_i \,.
\ee

In the case $\mu=0$ discussed in Sec.~\ref{sec:2.5}, the projectability
condition has the exact same effect as in the case $\mu \neq 0$
considered above.

\subsection{Interpretation and analysis of the tertiary constraints ---
conditions on the lapse \texorpdfstring{$N$}{N} or not?}
\label{sec:2.7}

Let us discuss the conditions \eqref{constraint_on_N_2nd} and
\eqref{constraint_on_N_generic}. 
In order to obtain real dynamics, the requirement $N \neq 0$ is
physically necessary, and if such a solution does not exist, the system
would have to be concluded as physically inconsistent. Such
inconsistency has been shown to exist in the usual non-projectable \HL
gravity with a generic potential \cite{Henneaux:2010}, where only the
solution $N=0$ exists when asymptotically flat spaces with appropriate
boundary conditions are considered (see
\cite{Charmousis:2009,Li:2009,Blas:2009,Kobakhidze:2009} for other
analyses). When the lapse is fixed to $N=0$, the Hamiltonian constraints
of these theories are generically second-class. The special cases of GR
and ultralocal theory are known to be consistent and possess first-class
Hamiltonian constraints. 
Other cases of the potential can at most have first-class Hamiltonian
constraints in some parts of space, but not everywhere
\cite{Farkas:2010}.
Imposing additional constraints may, at least in some cases, provide a
way to define a consistent theory. For example, the case of low-energy
effective potential of the form $R$ provides a consistent theory when
the additional constraint $\pi=0$ is imposed
\cite{Henneaux:2010,Pons:2010,Restuccia:2010}, and even to be equivalent
to GR in the gauge $\pi=0$. Problems related to the modification of the
Hamiltonian constraint of GR have also been studied in
Ref.~\cite{Barbour:2002,Kocharyan:2009}.

For asymptotically flat spaces we assume the following asymptotic
behaviour of the canonical variables in asymptotically flat coordinates
($C=\mr{constant}$):
\bea
N = C + O\left(\frac{1}{r}\right) ,&& N^i = O\left(\frac{1}{r}\right)
,\nn
g_{ij} = \delta_{ij} + O\left(\frac{1}{r}\right) ,&& \pi^{ij} =
O\left(\frac{1}{r^2}\right) ,\nn
B = 1 + O\left(\frac{1}{r}\right) ,&& \pi_B =
O\left(\frac{1}{r^2}\right) ,
\eea
which are an extension of the standard boundary conditions of GR
\cite{Regge:1974} that have been used in the usual \HL gravity
\cite{Henneaux:2010}. 
The auxiliary field $A$ is expected to behave as $\tildeR$ for obvious
reasons. 

The interpretation of the conditions \eqref{constraint_on_N_2nd} and
more generally \eqref{constraint_on_N_generic} is similar compared to
the analysis of the usual \HL gravity given in
Ref.~\cite{Henneaux:2010}. They are generically conditions on the lapse
$N$ and the only solution that satisfies the appropriate boundary
conditions in asymptotically flat space is $N=0$. Mathematically the
constraint $N \approx 0$ is perfectly acceptable, and setting the
Lagrange multiplier $\lambda_N=0$ ensures that this constraint is
preserved under dynamics. But as we have stated, this constraint ruins
the chance of having any meaningful gravitational dynamics. $N=0$
implies that the Hamiltonian vanishes in the gauge $N^i=0$, which means
that every function is a constant of motion. Moreover, it is unlikely
that the constraint \eqref{constraint_on_N_generic} could be satisfied
by introducing additional constraints that are both consistent under
dynamics and do not constrain the canonical variables too much. Such
tertiary 
constraints, together with the existing primary and secondary
constraints, would essentially have to imply $E_n \approx 0$ in
\eqref{PB_of_Phi0s_6th}--\eqref{constraint_on_N_generic}. We will
demonstrate this by considering the simple special case
\eqref{constraint_on_N}.

In the low-energy effective case \eqref{cLR_1st} with only one
nonvanishing coupling $\alpha_1$ the condition
\eqref{constraint_on_N_2nd} cannot be satisfied by introducing a simple
additional constraint like $\pi=0$ in the usual theory 
\cite{Henneaux:2010,Pons:2010,Farkas:2010,Restuccia:2010}. Instead we
could try to introduce more complex constraints to serve the same
purpose. In order to satisfy \eqref{constraint_on_N_2nd} the following
three constraints could be imposed:
\bea \label{extra_constraints}
\frac{1}{2}\tilde{E}_1^i &=& - \frac{2(\alpha_1+1)}{3}\nabla^i \pi 
+ 4(-\alpha_1+\mu) \left( \pi^{ij} - \frac{1}{3}g^{ij} \pi \right)
\frac{\nabla_j B}{B} \nn
&-& \frac{1-2\alpha_1-3\lambda}{3\mu} B \nabla^i \pi_B + \left(
\frac{3\lambda+\mu-2\alpha_1-1}{3\mu}+\alpha_1 \right) \nabla^i B \pi_B
\approx 0 \, .
\eea
It would require constraints like $\pi(\vect{x}) \approx 0$,
$\pi_B(\vect{x}) \approx 0$ and $B(\vect{x}) \approx \mr{constant}$
(i.e. ``projectable $B$'') to make the $\tilde{E}_1^i$ vanish due to
relatively simple constraints. But we can see from
(\ref{PB_g_ij_Phi0})--(\ref{PB_Phi0_piB}) that the preservation of
$\pi(\vect{x}) \approx 0$ and $\pi_B(\vect{x}) \approx 0$ under dynamics
would impose additional (quartic) constraints on the canonical
variables. These constraints most likely constrain the system too much,
leaving too little physical degrees of freedom for gravitational
dynamics. If we assume that only two quartic constraints would be needed
and that all these five constraints would be second-class and no more
constraints are needed to ensure their consistency, and that the
variables $A$ and $\pi_A$ can be eliminated (see Sec.~\ref{sec:2.4}), we
find that the (maximum) number of physical degrees of freedom would be
$3/2$ at each point. This is one degree of freedom less than in the
usual \HL 
gravity and one half degree of freedom less than in GR. Thus no more
than three constraints, e.g. \eqref{extra_constraints}, can be
introduced to satisfy the constraint \eqref{constraint_on_N} or
\eqref{constraint_on_N_2nd}. But the consistency of the three extra
constraints \eqref{extra_constraints} under dynamics is problematic ---
$\dot{\tilde{E}}_1^i$ is a long and complex expression that appears to
imply further (quartic) constraints, which would again lead to the lack
of degrees of freedom. In the more general case
\eqref{constraint_on_N_generic} it is practically certain that imposing
$E_n \approx 0$ constrains the system too much. A somewhat similar
problem has been discovered in the original \HL theory \cite{Li:2009}.

\section{Conclusion}

As a summary, there are serious problems with the physical consistency
of the non-projectable version of our theory, which will likely be
impossible to resolve with additional constraints. For the general
potential \eqref{cLR_general} the conclusion is similar compared to the
usual \HL gravity: since $N=0$ is required by the preservation of the
Hamiltonian constraints under time evolution, the theory is physically
inconsistent. The difference is that in our $F(R)$ gravity version, the
undesirable condition $N=0$ cannot be avoided consistently even for the
low-energy effective action. In the light of Ref.~\cite{Farkas:2010}, it
is very unlikely that there would exist a form of the action with
scaling dimension $z=3$ that could avoid this problem in the usual \HL
gravity. The same is expected to apply in our theory.

Thus, we can conclude that only the version of the modified $F(R)$ \HL
gravity with a projectable lapse, $N=N(t)$, is a practicable theory of
modified gravity. The non-projectable version of the theory is troubled
by a similar physical inconsistency as the usual non-projectable \HL
gravity.

It would be very interesting to understand whether some further
generalization of the \HL gravity could be defined consistently without
the projectability condition on the lapse --- perhaps even along the
lines of the general modified first-order \HL gravity of
Ref.~\cite{Carloni:2010}. It would require that the coefficients $E_n$
in \eqref{PB_of_Phi0s_6th}--\eqref{constraint_on_N_generic} take a very
special form, preferably vanish altogether. Another intriguing prospect
would be a generalization of the ``healthy extension'' of usual \HL
gravity~\cite{Blas:2010} to our modified $F(R)$ \HL gravity.

\section*{Acknowledgements}

The support of the Academy of Finland under the Projects No. 121720 
and 127626 is gratefully acknowledged. 
M. O. is supported by the Finnish Cultural Foundation.

\end{document}